\documentstyle[12pt]{article}

\newcommand{\be}{\begin{equation}}
\newcommand{\ee}{\end{equation}}
\newcommand{\ba}{\begin{eqnarray}}
\newcommand{\ea}{\end{eqnarray}}

\newcommand{\tr}{\rm tr}

\begin{document}
\hoffset=-.4truein\voffset=-0.5truein
\setlength{\textheight}{8.5 in}
\begin{titlepage}
\hfill{LPTENS 07-17}
\begin{center}

\vskip 0.6 in
{\large   Vertices from replica
in a random matrix theory}
\vskip .6 in
\begin{center}
{\bf E. Br\'ezin$^{a)}$}{\it and} {\bf S. Hikami$^{b)}$}
\end{center}
\vskip 5mm
\begin{center}
{$^{a)}$ Laboratoire de Physique
Th\'eorique, Ecole Normale Sup\'erieure}\\ {24 rue Lhomond 75231, Paris
Cedex
05, France. e-mail: brezin@lpt.ens.fr{\footnote{\it
Unit\'e Mixte de Recherche 8549 du Centre National de la
Recherche Scientifique et de l'\'Ecole Normale Sup\'erieure.
} }}\\
{$^{b)}$ Department of Basic Sciences,
} {University of Tokyo,
Meguro-ku, Komaba, Tokyo 153, Japan. e-mail:hikami@dice.c.u-tokyo.ac.jp}\\
\end{center}
\vskip 3mm 
{\bf Abstract}
\end{center}
Kontsevitch's work on Airy matrix integrals has led to explicit results for the
intersection numbers of the moduli space of curves. In a subsequent work Okounkov rederived these results from the edge behavior of a Gaussian matrix integral. In our work we consider  the correlation functions of  vertices in  a Gaussian random matrix theory , with an external matrix source. 
We deal with operator products of the form $< \prod_{i=1}^n \frac{1}{N}{\tr} M^{k_i}>$,
 in a $\frac{1}{N}$ expansion.  For  large values 
of the powers $k_i$, in an appropriate scaling limit relating large $k$'s to large
$N$,  universal scaling functions are derived. Furthermore we show that the replica method applied to characteristic polynomials of the random matrices, together with a duality exchanging N and the number of points, allows one to recover  Kontsevich's results on the intersection numbers, through a simple saddle-point analysis.
\end{titlepage}
\vskip 3mm
 
\section{Introduction}

Random matrix theory (RMT) has been applied to  many physical problems, and also to
mathematical subjects such as   the distribution of zeros of Riemann zeta
function or  combinatorial problems
and it has led to several meaningful results \cite{Brezinetal}.  It also plays an essential
role in the theory of random surfaces and for  string theory.
Several kinds of correlation functions in  random matrix theory have
been studied.  In  previous papers, we have studied the correlation function of the
eigenvalues
\cite{BHc}, and  the correlations of  the characteristic polynomials\cite{BHe,BHg},
for which we have derived  explicit integral representations.

In this article, we consider the correlation functions of  vertices on the basis of
previously derived  integral representations.  The diagrammatic
representation of the vertex $\langle{\tr} M^k \rangle$, where $M$ is a random matrix, is
obtained through Wick's theorem, by the pairings of $k$-legs,  each leg carrying the
two indices ($i,j$) of the matrix element
$M_{ij}$.
For $N\times N$ matrices, the two indices run from 1 to N :  $i,j=1,...,N$.

We restrict ourselves in this article to complex Hermitian
random matrices. The distribution function for $M$ is Gaussian with an
external matrix source
$A$.
\be\label{PA}
P_A(M) = \frac{1}{Z_A} e^{-\frac{N}{2}{\tr} M^2 - N {\tr} M A}
\ee
When one sets $A=0$, it reduces to the usual Gaussian unitary ensemble (GUE).

The correlation functions for the vertices $V(k_1,...,k_n)$ are defined as
\be
V(k_1,...,k_n) = \frac{1}{N^n}<{\tr} M^{k_1} {\tr} M^{k_2} \cdots {\tr} M^{k_n}>
\ee
The normalization is chosen so that they have a finite large-$N$ limit. These functions are
closely related to the Fourier transform of the correlation functions of the eigenvalues,

\be
U(t_1,...,t_n) = \int_{-\infty}^\infty e^{i \sum t_i
\lambda_i}R_n(\lambda_1,...\lambda_n) \prod_1^N d\lambda_i 
\ee
where the correlation function of the eigenvalues is
\be
R_n(\lambda_1,...,\lambda_n) = < \prod_{i=1}^n \frac{1}{N}{\tr} \delta (\lambda_i-M)>
\ee

Indeed 
\be
U(t_1,...,t_n) = <\frac{1}{N^n}\prod_{i=1}^n {\tr} e^{i  t_i M}>
\ee
 are generating functions of the $V(k)$ since
\be\label{gen}
U(t_1,...,t_n) = \sum_{k_i=0}^\infty < {\tr} M^{k_1}{\tr} M^{k_2} \cdots {\tr} M^{k_n}> 
\frac{(it_1)^{k_1}\cdots (it_n)^{k_n}}{k_1!k_2!\cdots k_n! N^n}
\ee

When the distribution of the random matrix is  Gaussian, the average
of the vertices gives the numbers of pairwise gluing  of the legs of the vertex operators.
The dual cells of these vertices are polygons,
whose edges are pairwise glued. We thereby generate orientable surfaces, 
which are discretized
Riemann surfaces of  given genus.
 
Okounkov and Pandharipande \cite{OP,O1} 
have shown that the intersection numbers,  computed by Kontsevich \cite{Kontsevich},
may be obtained by taking
a simultaneous large $N$ and large $k_i$ limit. Furthermore the correlation functions of
these vertices are  interesting, since they give  universal numbers in the large N
limit. 
 
We  have investigated in an earlier work  the F.T. of the n-point correlation function 
$U(t_1,...t_n)$ for the GUE, and found
a simple contour integral representation valid even for finite $N$ \cite{BHb,BHc}.
 
 In this article, we extend this integral representation to the vertex correlations
$V(k_1,.. k_n)$, and examine the scaling region for large $k_i$ and large N. 
In this integral representation, the asymptotic evaluation by the saddle-point method
requires a careful examination to deal with pole terms. This leads to a practical way
to compute  intersection numbers which we discuss  in detail.
We also show   that the F.T. of the correlation functions (C.F.) of GUE
near the edge point of the support of the asymptotic spectrum, is  equivalent to
Kontsevich's Airy matrix model ; the identification is based on 
the  replica method and over a duality for computing averages of characteristic
polynomials.

The article  is organized as follows.

In section 2, we consider the F.T. of the one point correlation function 
at a bulk generic point in the large $N$ limit. This is done by  a contour
integral representation, and we obtain the behavior of $<{\tr} M^{2k}>$ 
when $N$ and $k$ are large. 
We show that in this limit, one recovers the behavior of
 the one point function near the
edge point of the spectrum.

In section 3, we consider the correlation function of two vertices.

In section 4, we investigate the correlations of the n-vertices.

In section 5, 
we introduce a replica method, relying on averages of characteristic polynomials. This,
together with a duality, allows us to make  connexion with the  Kontsevich
model, recovering  thereby generating functions for the
intersection  numbers.

In section 6, we present a short summary.

\section{One point correlation function}
\vskip 2mm

The correlation function $R_n(\lambda_1,...,\lambda_n)$ defined by
\be
R_n(\lambda_1,...,\lambda_n) = < \prod_{i=1}^n \frac{1}{N}{\tr} (\delta(\lambda_i - M))>.
\ee
is thus equal to 
\be
R_n(\lambda_1,...,\lambda_n) = < \prod_{i=1}^n \frac{1}{N}{\tr} (\int_{-\infty}^\infty  
\frac{dt}{2\pi} e^{-i t_i(\lambda_i - M)})>.
\ee

The Fourier transform $U(t_1,...,t_n)$ of $R_n(\lambda_1,...,\lambda_n)$ is thus given by
\ba
U(t_1,...,t_n) &=& \int_{-\infty}^\infty \prod_{i=1}^N d \lambda_i e^{i\sum _i t_i \lambda_i}
R_n(\lambda_1,...,\lambda_n)\\
&=& < \frac{1}{N^n}\prod_{i=1}^n {\tr} e^{i  t_i M}>  .
\ea
Note that $U(t_1,...,t_n)$ is normalized to  one when all $t_i=0$.
The function $U(t_1,...,t_n)$ is the generating function of the correlation
$V(t_1,...,t_n)$ as shown in (\ref{gen}).

These F.T. of the correlation functions $U(t_1,...,t_n)$ were investigated in our earlier 
study of the kernels for the correlation functions  \cite{BHa,BHb} ; there we had derived
an exact integral representation for these correlation functions. We consider  
the probability distribution
$$P_A(M) = \frac{1}{Z_A} e^{-\frac{N}{2}{\tr} M^2 - N {\tr} M A}$$
then one finds the exact result  
\ba\label{Unpoint}
U(t_1,...,t_n) &=& \frac{1}{(t_1 \cdots t_n)}e^{- \frac{1}{2N}(t_1^2 + \cdots + t_n^2)}
\nonumber\\
&\times& \oint \frac{du_1\cdots du_n}{(2\pi i)^n}
e^{i  \sum_p t_p u_p}\prod_{p=1}^n \prod_{\gamma=1}^N ( 1+ \frac{it_p}{N(u-a_\gamma)})
\nonumber\\
&\times& \prod_{p<q}^n \frac{[u_p - u_q + \frac{i}{N}(t_p - t_q)](u_p - u_q)}{(u_p - u_q + \frac{i}{N}
 t_p)(u_p
- u_q - \frac{i}{N} t_q)}
\ea
where the integration contours circle around the eigenvalues $a_{\gamma}$ , $(\gamma = 1,
\cdots N)$, of the source matrix $A$ in the anticlockwise direction.

For the one-point function
\be\label{UUt}
U(t) = <\frac{1}{N} {\tr} \hskip 1mm e^{i t M} >
\ee
the exact integral representation for finite N is thus \cite{BHa}
\be\label{Ut}
U(t) = \frac{1}{it} \oint \frac{du}{2 \pi i} \prod_{\gamma=1}^N (\frac{u - a_\gamma + \frac{it}{N}}
{u - a_\gamma}) e^{-\frac{t^2}{2N} + i t u} ,
\ee

which reduces,  for  the pure Gaussian model $a_\gamma=0$,   to 
\be\label{Uexact}
U(t) = \frac{1}{it} \oint \frac{du}{2 \pi i} (1+\frac{it}{Nu})^N e^{-\frac{t^2}{2N} + i t
u}.
\ee
For this sourceless GUE one obtains immediately in the large $N$ limit 
\be
U(t) = \frac{1}{it} \oint \frac{du}{2\pi i} e^{it(u + \frac{1}{u})} .
\ee

The generating function for Bessel functions
$J_j(x)$,
\ba\label{Bessel}
&&e^{it(u + \frac{1}{u})} = \sum_{j=-\infty}^\infty (iu)^j J_j(2t)\nonumber\\
&&J_{-j}(x) = (-1)^j J_j(x) ,
\ea
leads to
\be\label{Ut1}
U(t) 
= \frac{1}{t}J_1(2t) .
\ee

The semi-circle law for the density of states of the  GUE follows :

\ba\label{square}
  \rho(x) &=& \int_{-\infty}^\infty \frac{dt}{2\pi} U(t) e^{-i x t} \nonumber\\
&=& 0, \hskip 32mm (\frac{x}{2} \ge 1)\nonumber\\
&&  \frac{1}{\pi} \sqrt{1 - (\frac{x}{2})^2},\hskip 10mm (\frac{x}{2} \le 1) .
\ea

Returning now to the exact expression (\ref{Uexact}) for finite $N$ 
one finds
\ba\label{genericA}
U(t) &=& \sum_{k=0}^\infty \frac{(it)^{2k}}{N^k}
[\sum_{l=0}^k \frac{\Gamma(N)}{\Gamma(N-k+l)\Gamma(k-l+1)\Gamma(k-l+2)\Gamma(l+1)2^l}]
\nonumber\\
&=& 1 + \frac{(it)^2}{2} + \frac{1}{12}(1 + \frac{1}{2N^2}) (it)^4 + 
\frac{1}{144}(1 + \frac{2}{N^2})(it)^6 + \cdots\nonumber\\
\ea
There are no odd powers of $\frac{1}{N}$ 
 in this expansion, as is well-known for the
GUE case, for which the successive terms of the $1/N$-expansion are of the form
$1/N^{2g}$, where $g$ is the genus of the surface generated by the Wick contractions.  
From the relation between
$U(t)$ and
$<{\tr} M^{2k}>$, we obtain
\ba\label{exact1}
&&\frac{1}{N}<{\tr} M^{2k}> = \frac{(2k)!}{N^k}[\sum_{l=0}^k 
\frac{\Gamma(N)}{\Gamma(N-k+l)\Gamma(k-l+1)\Gamma(k-l+2)\Gamma(l+1)2^l}]\nonumber\\
&=& \frac{(2k)!}{N^k}[ \frac{(N-1)(N-2) \cdots (N-k)}{k! (k+1)!} + \frac{(N-1)(N-2)
\cdots (N-k+1)}{(k-1)!k! 2}\nonumber\\
&+& \frac{(N-1)(N-2) \cdots (N-k+2)}{(k-2)!(k-1)! 8} + \cdots]
\ea

This exact representation
leads to  (\ref{exact1}) the expansion  
\ba\label{trM}
&&\frac{1}{N}<{\tr} M^{2k}> = \frac{(2k)!}{k!(k+1)!}[ 1 + \frac{k(k-1)(k+1)}{12 N^2}
\nonumber\\ && + \frac{k(k+1)(k-1)(k-2)(k-3)(5 k - 2)}{1440 N^4} + O(\frac{1}{N^6})]
\ea

The large N limit is  the first term of the above expansion :
\be\label{Catalan}
\lim_{N\to \infty} \frac{1}{N}< {\tr} M^{2k} > = \frac{(2k)!}{k! (k+1)!}
\ee
 the  k-th Catalan numbers. For large $k$ this number behaves as
$\frac{1}{\sqrt{\pi}}\frac{1}{k^{\frac{3}{2}}}4^k $. 
Therefore the resolvent 
\ba
G(\lambda ) &=& < \frac{1}{N}{\tr} \frac{1}{\lambda - M} >\nonumber\\
&=& \frac{1}{\lambda} \sum_{k=0}^{\infty} \frac{1}{\lambda^{2k}} <{\tr} M^{2k} >
\ea
has 
a square root singularity at $\lambda_c^2= 4$.  This corresponds of course to the vanishing
of the asymptotic density of state (\ref{square}) as a square root at the edge.

Returning to the  $\frac{1}{N}$ expansion of $<\tr M^{2k}>$  one finds for large k
\ba
&&\frac{1}{N}<{\tr} M^{2k}> \sim \frac{1}{\sqrt{\pi}}\frac{1}{k^{\frac{3}{2}}}4^k (1 - 
\frac{21}{8 k} + O(\frac{1}{k^2}))\nonumber\\
&&\times [ 1 + \frac{k(k^2-1)}{12 N^2} +
\frac{k(k+1)(k-1)(k-2)(k-3)(5 k - 2)}{1440 N^4} + O(\frac{1}{N^6})] . \nonumber\\
\ea 
When $k$ is of the order $k\sim N^{\frac{2}{3}}$, the above
series exhibits a scaling behavior:

\be\label{sc}
\frac{1}{N}<{\tr} M^{2k}> \sim \frac{1}{\sqrt{\pi}k^{\frac{3}{2}}}4^k 
[ 1 + \frac{k^3}{12 N^2} +
\frac{k^6}{(12)^2 2!N^4} + O(\frac{1}{N^6})] .
\ee 
The power $4^k$ corresponds to the location of the edge of the support of the
asymptotic spectrum 
$(\lambda_c^2 = 4)$,  and it is not universal.
But the successive terms being powers of  $k^3/N^2$ is universal, since this feature is related to the square root vanishing of the density of states.
The  scaling function in this double limit of large N and large k 
when
$k$ behaves as 
$k\sim N^{2/3}$,
is also universal.
The above  coefficients  of the scaling function in (\ref{sc}) provide the intersection
numbers of the moduli of curves.
The universality  of the coefficients of the series in powers of  $k^3/N^2$ corresponds to
the F.T. near the edge of the spectrum $\lambda\sim\lambda_c$, for the universal Airy kernel.

In order to obtain the full scaling function of $k^3/N^2$, and not simply the first terms
of the expansion as in (\ref{sc}) ,  we now consider an exact integral representation for
$<{\tr} M^{2k}>$. From (\ref{gen})
\ba\label{dtdu}
\frac{1}{N}<{\tr} M^{2k}> &=& (2k)! \oint \frac{dt}{2\pi i} \frac{U(t)}{i^{2k}
t^{2k+1}}\nonumber\\ &=& \frac{(2k)!}{i(-1)^k}\oint \oint \frac{dt du}{(2\pi i)^2}
\frac{1}{t^{2k+2} }(1 + \frac{it}{Nu})^N
e^{-\frac{t^2}{2N} + i t u}
\ea

For large k, large N, we apply the saddle-point method to (\ref{dtdu}).
The integrand in (\ref{dtdu}) behaves as $e^{N \phi}$, with
\be
\phi =- 2k \ln t + \frac{it}{u} + i t u
\ee
The saddle point equations,
\ba
 &&\frac{\partial \phi}{\partial u} = it [ -\frac{1}{u^2} + 1] = 0\nonumber\\
 &&\frac{\partial \phi}{\partial t} = -\frac{2k}{t}+ i (\frac{1}{u} + u) =0
\ea
give as solutions  $u=\pm 1, t= \mp i k$.
Expanding around the saddle-point in a standard way one obtains in the large $k$, large
$N$ limit 
\be
\frac{1}{N}<\tr M^{2k}> \sim \frac{(2k)! e^{2k}}{k^{2k+1}}\frac{1}{4\pi k}e^{\frac{k^3}{12
N^2}}
\sim \frac{4^k}{\sqrt{\pi}k^{\frac{3}{2}}}e^{\frac{k^3}{12 N^2}}
\ee

We have thus obtained 
the scaling function 

\be\label{sc2}
f(\frac{k^3}{N}) = \exp[ \frac{k^3}{12 N}]
\ee
  in accordance  with the expansion found hereabove in (\ref{sc}).

Instead of  the bulk spectrum, we now consider the F.T. of the  one point correlation function
near  the edge point ($\lambda \sim \lambda_c$). We  denote by $\tilde U(t)$ 
the F.T. of the one point correlation function near the edge of the spectrum, to distinguish
it from the bulk  $U(t)$. We expect from the previous argument
that this
$\tilde U(t)$ becomes  $<\tr M^{2k}>$ if one puts $ t = - ik$.

To explore the vicinity of the edge, it is convenient to introduce a trivial 
 external matrix source whose eigenvalues are all
$a_{\gamma}=-1$ in (\ref{Ut}), and multiply
$e^{-it}$ in order to compensate for this uniform shift. Thereby the edge is now at the
origin and 
\ba\label{Uedge}
\tilde U(t) &=& \frac{e^{-it}}{i t} \oint \frac{du}{2 \pi i}
( 1 + \frac{it}{N(1+u)})^N 
e^{-\frac{t^2}{2N} + i t u}\nonumber\\
&=& \frac{e^{-it}}{it}\oint \frac{du}{2\pi i} e^{N ln( 1 + \frac{it}{N(1+u)})}
e^{-\frac{t^2}{2N} + i t u}
\ea
In the regime in which $t\sim N^{2/3}$ and $u\sim N^{-1/3}$ one may expand for small $u$ up
to order
$u^2$, and the contour integral becomes a saddle point Gaussian integral. Note that
the term $it u$ is cancelled in the exponent. We have
\ba\label{tildeU}
\tilde U(t) &=&  \frac{1}{it} e^{ \frac{(it)^3}{3 N^2}  } 
\int_{-\infty}^\infty \frac{du}{2\pi i} e^{ i t u^2 - \frac{1}{N}
t^2 u}\nonumber\\
&=& \frac{1}{2\sqrt{\pi}(it)^{\frac{3}{2}}}e^{\frac{(it)^3}{12N^2}}
\ea
We thus recover the scaling function $f= \exp[ k^3/12N^2]$, when we put $t= -ik$
in (\ref{tildeU}).The difference between $\frac{1}{N}<{\tr} M^{2k}>$ and $\tilde
U(t)$ when  we replace  $it$ by $k$, is only the prefactor $\frac{1}{2} 4^{k}$, .
We have thus shown that the scaling function of (\ref{sc2}) in $<{\tr} M^{2k}>$ for $k\sim
N^{\frac{2}{3}}$,namely $\exp[ \frac{k^3}{12 N^2}]$, may also be obtained
from the F.T. of the one point correlation function near the edge  by setting
$t = -ik$.

Changing the prefactor by multiplying by $1/t^{1/2}$, and defining $x= - i
t/2^{1/3}$,
 we obtain the  generating
function,
\ba\label{Fx}
F(x) &=& \frac{1}{x^2} e^{\frac{x^3}{24 N^2}}\nonumber\\
&=& \frac{1}{x^2} + \frac{x}{24 N^2} + \frac{x^4}{(24)^2 2! N^4} + \cdots\nonumber\\
&=& \sum_{g=0}^\infty <\tau_{3g - 2}> \frac{1}{N^{2g}} x^{3g - 2}.
\ea

The numbers $<\tau_{k}>$ in this expansion coincide with the intersection number of the
moduli of curves,  as we will be justified below
by the replica method.

From the expansion  (\ref{Fx}), we obtain $<\tau_j>$ as
\be\label{tau}
  <\tau_{3g - 2} >_{g} = \frac{1}{(24)^g g!} \hskip 4mm (g=0,1,2,...).
\ee
These numbers agree with the values of the intersection numbers
computed earlier by Kontsevitch and Witten \cite{Kontsevich,Witten}. For  $<\tau_0>$, we need a special
consideration in (\ref{Fx}).  If we put $g= \frac{2}{3}$, we get $<\tau_0>$. However by
definition  $<\frac{1}{N} {\tr} M^{2k}>=1$ for $k=0$. Therefore we define 
$<\tau_0>_{g=0}=1$ instead of $<\tau_{-2}>_{g=0}=1$ for comparison with the intersection
numbers. We will discuss these intersection numbers later.

We have used here the integral representation to derive  $\tilde U(t)$. Since this is
related to the edge problem, we could have used instead the Airy kernel
$K_A(\lambda,\lambda)$,
\ba
K_A(\lambda,\mu) &=& 
\frac{A_i^{\prime}(\lambda) A_i(\mu) - A_i(\lambda) A_i^{\prime}(\mu)}{\lambda - \mu}\nonumber\\
&=& \int_0^\infty A_i(\lambda + z) A_i(\mu + z) dz
\ea
where
the Airy function $A_i(x)$ is given by
\be
A_i(x) = \frac{1}{2\pi} \int_{-\infty}^\infty e^{\frac{i}{3} \xi^3 + i x \xi} d \xi .
\ee
Near the edge of the spectrum $\lambda=2$, the density of state is given by
$K_A(\lambda,\lambda)$  in the appropriate large N scaling.
Let us verify that one can recover  the previous result from there :
\ba\label{result}
\tilde U(t) &=& \int_{-\infty}^\infty d\lambda e^{i t \lambda} \int_0^\infty dz
A_i^2(\lambda + z) \nonumber\\
&=& \frac{1}{(2\pi)^2} \int_0^\infty dz 
\int_{-\infty}^\infty e^{i t \lambda} \int_{-\infty}^\infty
e^{\frac{i}{3}\xi^3 + i \xi (\lambda + z)} d \xi \int_{-\infty}^\infty
e^{\frac{i}{3}\eta^3 + i (-t + z)\eta }d\eta
\nonumber\\
&=& \frac{1}{2\pi}
\int_0^\infty dz \int_{-\infty}^\infty d\xi \int_{-\infty}^\infty d\eta
\delta (\xi + \eta + t) e^{\frac{i}{3}\xi^3 + \frac{i}{3}\eta^3 + i z (\xi + \eta)}
\nonumber\\
&=& \int_0^\infty dz \frac{1}{2\sqrt{i \pi t}}e^{\frac{(it)^3}{12} - i z t}
\nonumber\\
&=& \frac{1}{2\sqrt{\pi}}\frac{1}{(it)^{3/2}}e^{\frac{1}{12}(it)^3}
\ea
which coincides as expected with (\ref{tildeU}).
\vskip 2mm

\vskip 2mm
\section{  Two-point correlation function}
\vskip 3mm

In the case of the two-point correlation function, we have
\be
R_2(\lambda_1,\lambda_2) = <\frac{1}{N}{\tr} \delta(\lambda_1 - M) \frac{1}{N}{\tr} \delta
(
\lambda_2 - M)>
\ee
and the F.T. of $R_2(\lambda_1,\lambda_2)$ is
\be\label{Utr}
U(t_1,t_2) = < \frac{1}{N}{\tr} e^{i  t_1 M}\frac{1}{N}{\tr} e^{i  t_2 M}>
\ee
This correlation function has been obtained in closed form, for finite $N$,  with the help
of the  HarishChandra-Itzykson-Zuber integral, in \cite{BHb}
\ba\label{alpha1}
U(t_1,t_2)&=& \sum_{\alpha_1,\alpha_2}
\frac{\prod_{i<j} [ a_i - a_j + \frac{i}{N}t_1(\delta_{i,\alpha_1}-\delta_{j,\alpha_1})
+ \frac{i}{N}t_2(\delta_{i,\alpha_2} - \delta_{j,\alpha_2})]}{ 
\prod_{i<j} (a_i-a_j)}\nonumber\\
&\times& e^{it_1a_{\alpha_1} +  i t_2 a_{\alpha_2} -\frac{1}{2N}t_1^2 - \frac{1}{2N}
t_2^2 - \frac{1}{N}t_1 t_2 \delta_{\alpha_1,\alpha_2}}
\ea
This sum is then divided in two parts; $\alpha_1=\alpha_2$ and $\alpha_1\ne \alpha_2$.
The first part gives 
\be
U^{I}(t_1,t_2)= \sum_{\alpha_1}\prod_{i<j} \frac{[a_i - a_j + \frac{i}{N}(t_1+t_2)
(\delta_{i,\alpha_1} -
\delta_{j,\alpha_1})]}{(a_i-a_j)} e^{-\frac{1}{2N}(t_1+t_2)^2 + i (t_1+t_2)a_{\alpha_1}}
\ee
This may be  expressed as the contour integral 
\be
U^{I}(t_1,t_2) = \frac{1}{i(t_1+t_2)} \oint \frac{du}{2\pi i} \prod_{\gamma=1}^N
[ 1 + \frac{i(t_1+t_2)}{N(u-a_\gamma)}]
e^{i(t_1+t_2)u + \frac{1}{2N}((it_1)^2+(it_2)^2)}
\ee
which is nothing but $U(t_1+t_2)$ from (\ref{Ut}).

The second term is expressed by the double contour integral
\ba\label{U12}
 U^{II}(t_1,t_2) &=& e^{-\frac{1}{2N}(t_1^2+t_2^2)}\oint \frac{du_1 du_2}{(2 \pi i)^2}
e^{i  t_1 u_1 + i  t_2 u_2}\prod_{\gamma=1}^N ( 1 + \frac{it_1}{N(u_1-a_\gamma)})
(1 + \frac{i t_2}{N(u_2-a_\gamma)})\nonumber\\
&&\times \frac{1}{t_1t_2}\frac{(u_1-u_2 + \frac{1}{N}(it_1 - it_2))
(u_1 - u_2)}{(u_1 -u_2 + \frac{i}{N}t_1)(u_1-u_2 - \frac{i}{N}t_2)}
\ea

Noting that
\be
1 - \frac{t_1t_2}{N^2(u_1-u_2 + \frac{i}{N}t_1)(u_1 - u_2 - \frac{1}{N}it_2)} =
\frac{(u_1-u_2 + \frac{1}{N}(it_1 - it_2))
(u_1 - u_2)}{(u_1 -u_2 + \frac{i}{N}t_1)(u_1-u_2 - \frac{i}{N}t_2)}
\ee
we find $U^{II}$ is a sum of a disconnected part and a connected part.

Therefore, we find the connected part of $U^{II}(t_1,t_2)$ 
as the contour integral
\ba
U_c^{II}(t_1,t_2) &=& - e^{\frac{1}{2N}((it_1)^2 + (it_2)^2)}
\oint \frac{du_1du_2}{(2\pi i)^2}e^{it_1 u_1+it_2 u_2} ( 1 + \frac{it_1}{Nu_1})^N
(1+ \frac{it_2}{Nu_2})^N\nonumber\\
&\times& \frac{1}{N^2(u_1-u_2+ \frac{it_1}{N})(u_1-u_2 - \frac{it_2}{N})}
\ea
We have set here all the $a_\gamma=0$ to deal with the pure GUE. 
The contour is around $u_1,u_2=0$. This $U_c(t_1,t_2)$ may be expanded in powers of $t_1$ and $t_2$.
Together with the exponential factor, it yields the expansion 
\ba\label{U2t1t2}
&&U_c^{II}(t_1,t_2) = \frac{1}{N}[ 1 - \frac{1}{2}(t_1+t_2)^2 + \frac{1}{N} t_1 t_2  + \frac{1}{12}(1+ \frac{1}{2N^2})
(t_1^4+t_2^4)\nonumber\\
&& + (\frac{1}{3}-\frac{1}{2N} + \frac{1}{6N^2}) (t_1^3 t_2 + t_1 t_2^3)
+ (\frac{1}{2} - \frac{1}{2N} + \frac{1}{4N^2}) t_1^2 t_2^2+ O(t^6)]\nonumber\\
\ea
One may wonder why we have obtained odd powers of $\frac{1}{N}$ in this expression but
one can
check that combined  with $U(t_1+t_2)$, the odd power in $\frac{1}{N}$ of
$ U_c^{II}(t_1,t_2)$ cancels.

If we compute the  integral representation of $U_c^{II}(t_1,t_2)$ by deforming the
contours to  collect the contributions of the poles at
$u_1=\infty$ and
$u_2=0$, instead of $u_1=0,u_2=0$, we obtain
\be\label{U2t1t2inf}
U_c(t_1,t_2)_{(\infty,0)} = \frac{1}{N}[t_1 t_2 - \frac{1}{2}t_1^2 t_2^2 - \frac{1}{2}(t_1^3 t_2 + 
t_1 t_2^3) + O(t^6)]
\ee
From the expression of $U(t)$ and $U_c^{II}(t_1,t_2)$ found in (\ref{genericA}) and
(\ref{U2t1t2}), we have 
\be
\frac{1}{N}U(t_1+t_2) - U_c^{II}(t_1,t_2) = -\frac{1}{N}[t_1 t_2 -\frac{1}{2}(t_1^3 t_2+ t_1 t_2^3)
-\frac{1}{2}t_1^2 t_2^2 + O(t^6)]
\ee
which is indeed equal to $-U_c(t_1,t_2)_{(\infty,0)} $ in (\ref{U2t1t2inf}) 
and justifies  
 the deformation of the contour integration in order to collect the residue of the pole at
$u_1=\infty$.

Let us first consider  the connected part of the two-point correlation function in the
large N limit.

\be
U_{c}(t_1,t_2) = - \frac{1}{N^2}\oint \frac{du_1 du_2}{(2\pi i)^2}
e^{it_1 u_1 + i t_2 u_2 + \frac{it_1}{u_1} + \frac{it_2}{u_2}}\frac{1}{u_1^2}[
\sum_{n=0}^\infty (\frac{u_2}{u_1})^n]^2
\ee
Using the Bessel function formula (\ref{Bessel}),
we find from the residues  at $u_1=\infty$ and $u_2=0$,
\be
U_{c}(t_1,t_2) = \frac{1}{N^2}\sum_{l=0}^\infty (-1)^l (l+1) J_{l+1}(2 t_1)J_{l+1}(2 t_2)
\ee

Noting that the F.T. of the Bessel functions $J_{l+1}(2t)$ is
\ba
\int_{-\infty}^\infty dt J_{l+1}(2t) e^{i t \lambda} 
&=& 2 (4 - \lambda^2)^{-\frac{1}{2}} \cos [(l+1)\phi]\hskip 5mm ((l+1)\hskip 2mm is \hskip 2mm even)
\nonumber\\
\int_{-\infty}^\infty dt J_{l+1}(2t) e^{i t \lambda} 
&=& 2 i (4 - \lambda^2)^{-\frac{1}{2}} \sin [(l+1)\phi]\hskip 5mm ((l+1)\hskip 2mm is \hskip 2mm odd)
\nonumber\\
\sin \phi &=& \frac{\lambda}{2}.
\ea 
we obtain
by inverse F.T.
\ba
R_2(\lambda_1,\lambda_2) &=& \frac{1}{N^2 (2\pi i)^2}\int dt_1 dt_2 e^{i \lambda_1 t_1 + i \lambda_2 t_2}
\sum_{l=0}^\infty (-1)^l(l+1) J_{l+1}(2 t_1)J_{l+1}(2 t_2)\nonumber\\
&=& \frac{1}{N^2 (2\pi)^2} \sum_{l=0}^\infty (-1)^l (l+1)[
e^{i(l+1)\phi_1} + (-1)^{l+1}e^{-i(l+1)\phi_1}]\nonumber\\
&&\times [ e^{i(l+1)\phi_2}+(-1)^{l+1}e^{-i(l+1)\phi_2}]
\frac{1}{\sqrt{(4-\lambda_1^2)(4-\lambda_2^2)}}
\nonumber\\
&=&
 - \frac{1}{2 N^2 \pi^2}
\frac{1}{(\lambda_1 - \lambda_2)^2} 
\frac{4 - \lambda_1 \lambda_2}{\sqrt{(4 -\lambda_1^2)(4 - \lambda_2^2)}}.
\ea
This result agrees with the earlier derivation of   \cite{BZ}.

We now consider the integral representation for $<{\tr} M^{k_1}{\tr} M^{k_2}>$.
We have
\be
\frac{1}{N^2}<{\tr} M^{2k_1}{\tr} M^{2k_2}>_c=
\frac{(2k_1)!(2k_2)!}{(-1)^{k_1+k_2}}
 \oint \oint \frac{dt_1 dt_2}{(2\pi i)^2}\frac{U_c(t_1,t_2)}{t_1^{2k_1+1}
t_2^{2k_2+1}}
\ee
where $U_c(t_1,t_2) = U_c^{II}(t_1,t_2) - U(t_1+t_2)$.
\ba \label{trM2}
\frac{< {\tr} M^{2k_1}{\tr} M^{2k_2}>_c^{II}}{N^2} &=& - \frac{(2k_1)!(2k_2)!}{N^2
(-1)^{k_1+k_2}}
\oint \frac{dt_1 dt_2}{(2\pi i)^2}
\oint \frac{du_1 du_2}{(2\pi i)^2} \frac{e^{-\frac{1}{2N}(t_1^2+t_2^2) + i t_1 u_1 + i t_2 u_2}}{
t_1^{2k_1+1} t_2^{2k_2+1}}\nonumber\\
&&\times \frac{(1 + \frac{it_1}{Nu_1})^N (1 + \frac{it_2}{N u_2})^N}{(u_1 -u_2+\frac{it_1}{N})(u_1-u_2 - \frac{it_2}{N})}
\ea

We are interested in the large N and large $k_1,k_2$ behavior, but in the region in which the $k_i$ are of order $N^{2/3}$.
As for  the previous calculation of  \\ $\frac{1}{N}<{\tr} M^{2k}>$, after exponentiation , we find again that the saddle points are $t_{1c}= - i k_1$ and
$t_{2c}=-i k_2$ with $u_1=u_2=1$ in (\ref{trM2}). Then we
expand
$t_1$ and $t_2$ near the saddle-points
\ba
&&t_1 = - ik_1 (1+ v_1)\nonumber\\
&&t_2 = - ik_2 (1+v_2)
\ea
and expand for  $v_1,v_2$ small.  The integrations over $v_1,v_2$
become Gaussian, and they are equivalent to the replacement of $t_i$ by their 
saddle point values $- ik_i$ (i=1,2).
Therefore we have for large k and N, after the shift $u_i\to 1+ u_i$,
\ba\label{I2}
 &&\frac{1}{N^2}<{\tr} M^{2k_1}{\tr} M^{2k_2}>_c^{II}\nonumber\\
&&=
- C(k_1,k_2)e^{\frac{1}{3N^2}(k_1^3+k_2^3)}\int_{-\infty}^\infty \frac{du_1 du_2}{(2\pi i)^2} 
\frac{e^{k_1u_1^2 + \frac{k_1^2}{N}u_1 + k_2 u_2^2 + \frac{k_2^2}{N}u_2}}{
(u_1- u_2 + \frac{k_1}{N})(u_1-u_2-\frac{k_2}{N})}
\ea
where the constant $C(k_1,k_2)$ is
\ba
C(k_1,k_2) &=& \frac{(2k_1)!(2k_2)!e^{2k_1+2k_2}}{(-1)^{k_1+k_2}\sqrt{k_1 k_2}
 k_1^{2k_1}k_2^{2k_2}}\nonumber\\
&\sim& \frac{1}{4^{2k_1+2k_2} 4\pi k_1 k_2}
\ea

The integration in (\ref{I2}) requires a careful examination of the pole terms.
Since the denominator may vanish,  we use 
\be\label{principal}
\frac{1}{u_1 - u_2 + \frac{k_1}{N} - i\epsilon} = P(\frac{1}{u_1 -u_2 + \frac{k_1}{N}} ) 
+ i \pi \delta(u_1-u_2 + \frac{k_1}{N})
\ee
 The $\delta$ function contibution is nothing but $\frac{1}{N^2}<\tr M^{2k_1+2k_2}>$.
The principal part is evaluated by writing the denominator as
\be\label{denomi}
\frac{1}{u_1-u_2 + \frac{k_1}{N}} = -i \int_0^\infty d\alpha e^{i(u_1-u_2 + \frac{k_1}{N})\alpha}
\hskip 5mm (Im k_1 > 0)
\ee

The integration in (\ref{I2}) becomes
\ba\label{I2a}
&&I_2 = \int_{-\infty}^\infty \frac{du_1 du_2}{(2\pi i)^2}
\int_0^\infty d\alpha d\beta 
e^{\sum_{i=1}^2 (k_i u_i^2 + \frac{k_i^2}{N}u_i) + i \alpha (u_1-u_2 +\frac{k_1}{N}) 
+ i\beta (u_1 - u_2 - \frac{k_2}{N})}\nonumber\\
&&=  \frac{1}{4 \pi \sqrt{k_1 k_2}}\int_0^\infty d\alpha d\beta
e^{-k_1(\frac{k_1}{2N} + \frac{i(\alpha+\beta)}{2k_1})^2 - k_2 (\frac{k_2}{2N} -
 \frac{i (\alpha + \beta)}{2k_2})^2
+ \frac{i\alpha k_1}{N} - \frac{i\beta k_2}{N}}
\ea
In this representation we have  assumed that $Im k_1 > 0,Im k_2 <0$, but
after integration over $\alpha,\beta$, this condition becomes irrelevant.
We replace 
\be
\alpha+\beta = x,\hskip 4mm \beta=z
\ee
with
\be
0 < z < x
\ee
Thus we write $I_2$ as
\ba
&&I_2
= \frac{e^{-\frac{k_1^3+k_2^3}{4N^2}}}{4\pi \sqrt{k_1 k_2}} \int_0^\infty dx \int_0^x dz 
e^{\frac{k_1+k_2}{4k_1k_2}x^2 + \frac{i(k_1+k_2)}{2N}x -\frac{i(k_1+k_2)}{N}z}
\nonumber\\
&&= \frac{iN}{4 \pi \sqrt{k_1 k_2}(k_1+k_2)}e^{-\frac{k_1^3+k_2^3}{4N^2}
+\frac{k_1k_2(k_1+k_2)}{4N^2}}\nonumber\\
&&\times \int_{-\frac{ik_1k_2}{N}}^{\frac{ik_1k_2}{N}}dx 
e^{\frac{k_1+k_2}{4k_1k_2} x^2}
\ea
After mutiplication by $e^{\frac{k_1^3+k_2^3}{3N^2}}$, we obtain
\be
e^{\frac{1}{3N^2}(k_1^3+k_2^3)} I_2 = 
- \frac{\sqrt{k_1k_2}}{2\pi (k_1+k_2)}e^{\frac{(k_1+k_2)^3}{12N^2}}
\sum_{l=0}^\infty \frac{1}{l!(2l+1)4^l N^{2l}}(k_1k_2(k_1+k_2))^l
\ee
Finally
\be\label{generating0}
\frac{1}{N^2}<{\tr} M^{2k_1}{\tr} M^{2k_2}>_c = 
- C(k_1,k_2)\frac{1}{2\pi} \frac{\sqrt{k_1k_2}}{k_1+k_2}e^{\frac{1}{12}(k_1+k_2)^3}
\sum_{l=0}^\infty
\frac{[- k_1 k_2(k_1+k_2)]^l}{l!(2l+1)(4N^2)^l}
\ee
Note that in this evaluation, we have already subtracted the $\delta$-function term, 
when we have represented the pole terms in terms of integrals over $\alpha$ and
$\beta$.

For making contact with Kontsevitch normalization \cite{Kontsevich} ,
 we change  $k_1,k_2$ to $\frac{k_1}{2^{1/3}},\frac{k_2}{2^{1/3}}$, 
and  multiply a factor $\frac{2\pi}{\sqrt{k_1k_2}}$.
We obtain the intersection
numbers for two points (n=2) as an expansion of the error function.
\be\label{Fx1x2}
F(x_1,x_2) 
=  \frac{1}{x_1+x_2}e^{\frac{(x_1+x_2)^3}{24 N^2}}\sum_{l=0}^\infty
\frac{[x_1 x_2(x_1+x_2)]^l (-1)^l}{[8N^2]^l (2l+1) l!}
\ee
where we set $x_i = k_i$ within the appropriate factors.
This function $F(x_1,x_2)$ is a generating function of the intersection numbers.
If we expand it in powers of $x_1$ and $x_2$ 
\be\label{generating1}
F(x_1,x_2)=\sum_{l_1,l_2}<\tau_{l_1}\tau_{l_2}>_g \frac{ x_1^{l_1}x_2^{l_2}}{N^{2g}}
\ee
the coefficient $<\tau_{l_1}\tau_{l_2}>$ is the intersection number
for genus $g$ ;  the genus is specified by $3g - 1 = l_1 + l_2$, and then the coefficient
is then given by returning to (\ref{Fx1x2}).

As  seen previously for $<{\tr} M^{2k}>$, the asymptotic evaluation of \\$<{\tr}
M^{2k_1}{\tr} M^{2k_2}>$ is given by the Fourier transform $U(t_1,t_2)$ near the end
point.

From the above expression we obtain  coefficients, which are the
intersection numbers for n=2,
\ba
&&<\tau_2 \tau_0>_{g=1}=<\tau_0\tau_2>_{g=1} = \frac{1}{24}, \hskip 4mm
<\tau_5 \tau_0>_{g=2}= <\tau_0 \tau_5>_{g=2}=\frac{1}{(24)^2 2!},\nonumber\\
&&
\hskip 4mm <\tau_1^2>_{g=1} = \frac{1}{24}, \hskip 4mm
<\tau_4\tau_1>_{g=2}= \frac{1}{384},\hskip 4mm
<\tau_3\tau_2>_{g=2} = \frac{29}{5760}
\ea
These numbers agree with Witten's earlier results \cite{Witten}.
For $<\tau_0^2>_{g=0}$, we use the normalization, $<\tau_0^2>=1$, in analogy with $<\tau_0>=1$.

It is worth noticing that when we set $x_2=0$, i.e. $k_2=0$ in (\ref{generating0}),
we do obtain 
\be
F(x_1,0) = F(x_1) = \frac{1}{x_1}e^{\frac{x_1^3}{24 N^2}},
\hskip 4mm 
<\tau_{l_1}\tau_0> = <\tau_{l_1}> .
\ee
The above relation is the string equation, as will be explained later. 

\vskip 2mm
\section{ The n-point correlations}
\vskip 3mm
The  Fourier transform $U(t_1,....,t_n)$
is given by (\ref{Unpoint}).
Using Cauchy determinant formula, it is expressed as a determinant.
For the connected part, we take the  longest cyclic rings for the indices $p,q$.
\ba\label{Uc}
U_c(t_1,...,t_n) &=& \frac{1}{N^n} e^{- \frac{1}{2N}(t_1^2 + \cdots + t_n^2)}
\nonumber\\
&\times& \oint \frac{du_1\cdots du_n}{(2\pi i)^n}
e^{i  \sum_p t_p u_p}\prod_{p=1}^n \prod_{\gamma=1}^N ( 1+ \frac{it_p}{N(u_p-a_\gamma)})
\nonumber\\
&\times& \prod_{cycle}\frac{1}{\frac{i}{N}t_p + u_p - u_q},
\ea
where the last product is the maximal cycle for the indices (p,q).
For instance, in the case n=3, we have two longest cycles $(1\to 2 \to 3 \to 1)$ and
$(1\to 3 \to 2 \to 1)$,
\ba
&&\prod_{cycle}\frac{1}{\frac{i}{N}t_p + u_p - u_q}
= \frac{1}{(u_1-u_2 + \frac{it_1}{N})(u_2-u_3 + \frac{it_2}{N})(u_3-u_1+\frac{it_3}{N})}
\nonumber\\
&&+ \frac{1}{(u_1-u_3 + \frac{it_1}{N})(u_2-u_1 + \frac{it_2}{N})(u_3-u_2+\frac{it_3}{N})}
\ea
These two terms contribute to the connected part of $U(t_1,t_2,t_3)$ as
\ba\label{U3a}
&&U_c(t_1,t_2,t_3) = \frac{1}{N^3} e^{-\frac{1}{2N}(t_1^2+t_2^2 + t_3^2)}\nonumber\\
&&\times \oint \frac{du_1 du_2 du_3}{(2\pi i)^3}
e^{i\sum t_j u_j } [\frac{(1+ \frac{it_1}{Nu_1})^N (1+\frac{it_2}{Nu_2})^N
(1+\frac{it_3}{Nu_3})^N}{(u_1-u_2+\frac{it_1}{N})(u_2-u_3+\frac{it_2}{N})(u_3-u_1+ \frac{it_3}{N})}
\nonumber\\
&& + \frac{(1+ \frac{it_1}{Nu_1})^N (1+\frac{it_2}{Nu_2})^N
(1+\frac{it_3}{Nu_3})^N}{(u_1-u_3+\frac{it_1}{N})(u_2-u_1+\frac{it_2}{N})(u_3-u_2+ \frac{it_3}{N})}
]
\ea
We find that the two terms are identical, and they are symmetric polynomials of the $t_p$.

As we have seen, we have to add several terms to obtain the connected part of $U(t_1,...,t_n)$,
to deal with  the cases $\alpha_i=\alpha_j$ in the summation implied by (\ref{alpha1}). The evaluation
of this integral is an extenstion of the previous study of $I_2$.

For the n=3 case, we have in the large N limit, neglecting all $\frac{1}{N}$ terms,
\ba
I &=& \oint \frac{du_1 du_2 du_3}{(2\pi i)^3} \frac{e^{\sum_i i t_i u_i + \frac{i t_i}{u_i}}}{(u_1-u_2)
(u_2-u_3)(u_3-u_1)}\nonumber\\
&=&- \oint \frac{du_i}{(2\pi i)^3}\frac{1}{u_1^2 u_2}\sum_{l_1,l_2,l_3}(\frac{u_2}{u_1})^{l_1}
(\frac{u_3}{u_1})^{l_3}(\frac{u_3}{u_2})^{l_2}\nonumber\\
&& \times \sum_{s_1,s_2,s_3=-\infty}^\infty(iu_1)^{s_1}(iu_2)^{s_2}(iu_3)^{s_3}J_{s_1}(2 t_1)
J_{s_2}(2 t_2) J_{s_3}(2 t_3)\nonumber\\
&=& \sum_{l_i,m_j=0}^\infty  \frac{(-1)^{m_1+m_2+m_3} t_1^{2 m_1+l_1+l_3+1}
t_2^{2 m_2+ l_1-l_2}t_3^{2 m_3 + l_2 +l_3 + 1}}{m_1!(m_1+ 1 + l_1 + l_3)! m_2!(m_2+ l_1-l_2)!
m_3!(m_3+ l_2 + l_3 + 1)!}\nonumber\\
\ea
We consider the coefficients of $(-1)^{k_1+k_2+k_3}t_1^{2 k_1} t_2^{2 k_2} t_3^{2 k_3}$ of $I$, which is 
denoted by $I^{2k_1,2k_2,2k_3}$.
We put
\ba
m_1&=& k_1 - \frac{1}{2}-\frac{l_1}{2}-\frac{l_3}{2}\nonumber\\
m_2&=& k_2 - \frac{l_1}{2}+\frac{l_2}{2}\nonumber\\
m_3&=& k_3 - \frac{l_2}{2}-\frac{l_3}{2}-\frac{1}{2}
\ea
If $l_1$ is even, then $l_2$ is even and $l_3$ is odd. 
If $l_1$ is odd, then $l_2$ is odd and $l_3$ is even.
These two cases give the same result, and give a factor 2 for 
$l_1$  even.We change  $l_1\to 2 l_1$, $l_2\to 2 l_2$ and $l_3 \to 2 l_3+1$.

\ba
&&I^{2k_1,2k_2,2k_3}\nonumber\\
&&= \sum_{l_1,l_2,l_3=0}^\infty
\frac{1}{(k_1-l_1-l_3-1)!(k_1+l_1+l_3+1)!(k_2-l_1+l_2)!(k_2+l_1-l_2)!}\nonumber\\
&& \times \frac{1}{(k_3-l_2-l_3 -1)!(k_3+l_2+l_3+1)!}
\ea
This sum is expressed by the contour integration,
\ba\label{FF}
&&I^{2k_1,2k_2,2k_3}\nonumber\\
&&=\frac{1}{(2k_1)!(2k_2)!(2k_3)!}\oint \frac{dx dy dz}{(2\pi i)^3}
\frac{(1+x)^{2k_1}(1 + y)^{2k_2}(1+ z)^{2k_3}}{x^{k_1}y^{k_2+1}z^{k_3}(1- x y)(1-x z)(1- \frac{z}{y})}
\nonumber\\
&&=\frac{1}{(2k_1)!(2k_2)!(2k_3)!}\oint \frac{dx dy dz}{(2\pi i)^3}
\frac{(x+y)^{2k_1}(1+ y)^{2k_2}(1+ z y)^{2k_3}}{x^{k_1} y^{k_1+k_2+k_3+1}z^{k_3}(1-x)(1-z)(1- x z)}
\nonumber\\
\ea
where the contours are around $x=y=z=0$, and in the last line, we have made the change of variables
 $x\to \frac{x}{y}$ and $z\to z y$. When $k_1$, $k_2$ and $k_3$ are large, the saddle point
for this integrals are  $x_c=y_c=z_c=1$ ; however the denominator vanishes at this point.
Therefore, we first  deform the contours. 
The integral of (\ref{FF}) is invariant, except for a  sign, under the change of variables,
$x\to \frac{1}{x}$,$y\to \frac{1}{y}$ and $z\to \frac{1}{z}$, which transforms
 the contour around  $x=y=z=0$ into  a contour at 
$x=y=z=\infty$. By  Cauchy theorem, the sum of all residues has to 
 vanish if we  include the residues at infinity,
Thereby we obtain the following identity between the different contour integrals  for (\ref{FF}).
\be\label{id}
\oint_{x=y=z=0} F = -\frac{1}{2}[\oint_{x=1,y=0,z=0} + \oint_{z=1,x=0,y=0} + \oint_{x=\frac{1}{z},
y=0,z=0}] F
\ee
where $F$ is the integrand of (\ref{FF}). 
This identity is derived from the invariance, except for the overall  sign,  under the change
$x\to \frac{1}{x},y\to \frac{1}{y},z\to \frac{1}{z}$ for the expression (\ref{FF}).
We have 
\ba
&&\oint_{x=y=z=0} \frac{dx dy dz}{(2\pi i)^3}
\frac{(x+y)^{2k_1}(1+ y)^{2k_2}(1+ z y)^{2k_3}}{x^{k_1} y^{k_1+k_2+k_3+1}z^{k_3}(1-x)(1-z)(1- x z)}\nonumber\\
&&= - \oint_{x=y=z=\infty} \frac{dx dy dz}{(2\pi i)^3}
\frac{(x+y)^{2k_1}(1+ y)^{2k_2}(1+ z y)^{2k_3}}{x^{k_1} y^{k_1+k_2+k_3+1}z^{k_3}(1-x)(1-z)(1- x z)}\nonumber\\
\ea
Therefore, we obtain the  identity (\ref{id}) with the factor $1/2$.

The double pole $\frac{1}{(1-z)^2}$, which appears for the
contour integral around $x=1$, is transformed into a single pole by an integration 
by parts over $x$,
and the sigularity at $z=1$ is cancelled by the numerator.
Therefore  we may now use the saddle point at $z_c=1$, and obtain the large
$k_i$ behavior, 
\ba\label{asymptotic}
I^{2k_1,2k_2,2k_3}&=&
\frac{4^{k_1+k_2+k_3}}{4 \pi (2k_1)!(2k_2)!(2k_3)!  (k_1+k_2+k_3)}
[\sqrt{k_3(k_1+k_2)}\nonumber\\
&+&\sqrt{k_1(k_2+k_3)} -\sqrt{k_2(k_1+k_3)}]
\ea
In order to appreciate the asymptotic behavior of these integrals, we have compared the exact  value for the contour integral  (\ref{FF}), multiplied by a factor $(2k_1)!(2k_2)!(2k_3)!$,  to the large k estimates. For instance 
 for $k_1=30,k_2=15,k_3=50$ the exact integral is $7.4632 \times 10^{55}$, and the large $k_i$ asymptotic formula (\ref{asymptotic})  gives instead   $7.4864\times
10^{55}$. 

We have considered the region $u_1 \ge u_2 \ge u_3$. There are other regions $u_i\ge u_j \ge u_l$.
Adding  their contributions amounts to summing over  permutations of the  $k_1,k_2$ and $k_3$ ;   taking also into account
 the $I^{k_1,k_2,k_3}$ for $l_1$  odd, we obtain
\ba
&&I^{k_1,k_2,k_3}= \frac{4^{k_1+k_2+k_3}}{2\pi (2k_1)!(2k_2)!(2k_3)!
(k_1+k_2+k_3)}[\sqrt{k_3(k_1+k_2)}+\sqrt{k_1(k_2+k_3)}
\nonumber\\
&&+\sqrt{k_2(k_1+k_3)}]
\ea
This leads to,
\ba\label{3genus0}
&&\frac{1}{N^3}<{\tr} M^{2k_1}{\tr}M^{2k_2}{\tr}M^{2k_3}>_c\nonumber\\
&&= \frac{4^{k_1+k_2+k_3}}{2\pi (k_1+k_2+k_3)}[\sqrt{k_3(k_1+k_2)}+\sqrt{k_1(k_2+k_3)}
+ \sqrt{k_2(k_1+k_3)}]\nonumber\\
\ea
 
We have evaluated the leading term of order one in the large N limit of  the three-point correlation function. However, this leading term is cancelled when we consider connected correlation functions. 
Indeed  let us consider  the expansion of $U(t)$ and 
$U(t_1,t_2)$ : 
\ba
  U(t) &=&\frac{1}{N} \oint \frac{du}{2\pi i} \frac{N}{i t} e^{i t u} ( 1 + \frac{it}{N u})^N
e^{-\frac{t^2}{2 N}}\nonumber\\
&=& 1 - \frac{1}{2}t^2 + \frac{1}{12}t^4 + \frac{1}{24 N^2} t^4 + O(t^6)
\ea
\ba
&&U_c(t_1,t_2) = \frac{1}{N^2}e^{- \frac{1}{2N}(t_1^2+t_2^2)}\oint \frac{du_1 du_2}{(2 \pi i)^2}
\frac{e^{i t_1 u_1 + i t_2 u_2}(1 + \frac{ i t_1}{N u_1})^N ( 1+ \frac{i t_2}{N u_2})^N}{( u_1 - u_2 + \frac{i t_1}{N})(
u_2 - u_1+ \frac{i t_2}{N})}\nonumber\\
&&= \frac{1}{N}[ 1 - \frac{1}{2}(t_1 + t_2)^2 + \frac{1}{N} t_1 t_2 
 + \frac{1}{12}(t_1+t_2)^4 -\frac{1}{2 N} (t_1^3 t_2 + t_1^2 t_2^2 + t_1 t_2^3)\nonumber\\
&& + \frac{1}{6 N^2} (t_1^3 t^2 + t_1 t_2^3) + \frac{1}{4 N^2} t_1^2 t_2^2 + O(t^6)]
\ea
Then 
\ba\label{U123}
  &&U_c(t_1,t_2,t_3) = \frac{2}{N^3} e^{-\frac{1}{2 N} (t_1^2 + t_2^2 + t_3^2)}\nonumber\\
&& \times \oint \prod\frac{du_j}{(2\pi i)}\frac{e^{\sum i t_j u_j} \prod_{j=1}^3 
(1 + \frac{i t_j}{N u_j})^N}{
(u_1-u_2+ \frac{it_1}{N})(u_2-u_3 + \frac{i t_2}{N})(u_3 - u_1 + \frac{i t_3}{N})}\nonumber\\
&&= \frac{2}{N^2}[ 1 - \frac{1}{2}(t_1+t_2 +t_3)^2 + \frac{1}{12}(t_1+t_2 + t_3)^4
+ \frac{1}{N} (t_1 t_2 +t_1 t_3 +t_2 t_3)\nonumber\\
&&- \frac{1}{2 N}(t_1^2 t_2^2 + t_1^2 t_3^2 + t_2^2 t_3^2)
 - \frac{1}{2N}(t_1^3 t_2 + t_1^3 t_3 + t_1 t_2^3 + t_2^3 t^3 + t_1 t_3^3 + t_2 t_3^3)\nonumber\\
&&- \frac{2}{N} t_1 t_2 t_3 (t_1+t_2+t_3)+ \frac{1}{24 N^2} (t_1 + t_2 + t_3)^4\nonumber\\
&&+ \frac{1}{ 2 N^2} t_1 t_2 t_3 (t_1+t_2 + t_3)
+ O(t^6)]
\ea
Combining these expansions one obtains
\ba\label{cancel2}
&&U_c(t_1+t_2+t_3) - N U(t_1+t_2,t_3) - N U(t_1+t_3,t_2)  - N U(t_2+t_3,t_1)\nonumber\\
&&+ N^2 U_c(t_1,t_2,t_3)\nonumber\\
&& = \frac{1}{N^2}[ \frac{1}{8} (t_1^4+ t_2^4+ t_3^4) 
+ \frac{1}{4}(t_1^2 t_2^2 + t_1^2 t_3^2 + t_2^2 t_3^2)+ t_1 t_2 t_3 (t_1+t_2 + t_3)\nonumber\\
 &&+ \frac{1}{6}(t_1^3 t_2 + t_1^3 t_3 + t_2^3 t_1 + t_2^3 t_3 + t_3^2 t_1 + t_3^2 t_2)
 + O(t^6)]
\ea
It is order of $\frac{1}{N^2}$. Thus we see that the term, which we had considered in the large N limit, is cancelled 
by  the additional terms terms in (\ref{cancel2}). In other words, the odd-power  of $\frac{1}{N}$
are cancelled in the combination of (\ref{cancel2}).
We thus have to expand also the denominator in (\ref{U123}), and compute
the order $\frac{1}{N^4}$ , instead of the order $\frac{1}{N^3}$ in $U_c(t_1,t_2,t_3)$.

Noting that expressions for the large $N$
 limit of $\frac{1}{N}<{\tr } M^{2 k_1} {\tr} M^{2 k_2}>$ is given by 
\be
\frac{1}{N^2}<{\tr} M^{2 k_1}{\tr} M^{2 k_2}> \sim \frac{\sqrt{k_1 k_2}}{\pi N^2 (k_1+k_2)}4^{k_1+k_2},
\ee
we have 
\ba
&&\frac{1}{N^2}<{\tr} M^{2 k_1+ 2 k_2}{\tr} M^{2 k_3}> 
+ \frac{1}{N^2}<{\tr} M^{2 k_1+ 2 k_3}{\tr} M^{2 k_2}> \nonumber\\
&&+ \frac{1}{N^2}<{\tr} M^{2 k_2+ 2 k_3}{\tr} M^{2 k_1}> \nonumber\\
&&= - \frac{4^{k_1+k_2+k_3}}{\pi N^2 (k_1+k_2+k_3)}[ \sqrt{k_3(k_1+k_2)} + \nonumber\\ && \sqrt{k_2(k_1+k_3)}
+ \sqrt{k_1(k_2+k_3)}]  
\ea
This sum is identical to what we found  
in (\ref{3genus0}) for
$\frac{2}{N^2}<{\tr}M^{2 k_1} {\tr} M^{2 k_2} {\tr} M^{2 k_3} >$. 

To discuss the next order terms, we first derive a formula for the 
correlation functions of the vertices, which is applicable to the general 
case of n-point vertex correlations for arbitrary genus.
Let us use the following notation,
\be
[k_i k_j] = \frac{k_i+k_j}{2N}
\ee
Returning to the expression for the correlation function of n-point vertices 
in the large $k_i$ limit, one shifts  $u_i\to u_1-\frac{k_i}{2N}$, 
\ba
&& U(t_1,...,t_n)  
=  e^{\frac{1}{12 N^2}\sum k_j^3}\oint \frac{\prod du_i}{(2\pi i)^n}
\frac{e^{\sum i t_i k_i + \frac{i t_i}{u_i}}}{\prod_{i,j}(u_i - u_j + [k_i k_j])}\nonumber\\
&&= e^{\frac{1}{12 N^2}\sum k_i^3}\oint \frac{du_i}{(2\pi i)^n}
e^{\sum i t_i k_i + \frac{i t_i}{u_i}}
(\sum_{\nu_1} \frac{(- [k_1 k_2])^{\nu_1}}{(u_1 - u_2)^{\nu_1+1}})\cdots
(\sum_{\nu_n} \frac{(- [k_n k_1])^{\nu_n}}{(u_n-u_1)^{\nu_n+1}})\nonumber\\
\ea
From this representation one extracts $\frac{1}{N^n}<\prod_{j=1}^n {\tr } M^{2 k_j} >_c$ as coefficient of the
relevant power  of $t_i$. 

The above formula is applicable to the  n-point case for arbitrary genus.
For example, we consider  n=3, and  evaluate $\frac{1}{N^3}<\prod_{i=1}^3 {\tr}M^{2 k_i}>$.

We use the notation,
\be
C_l(\nu) = \frac{(l+\nu)!}{l! \nu!}
\ee
with $C_l(0)=1$.
Expanding 
\be
\frac{1}{(u_1-u_2)^{\nu+1}} = \frac{1}{u_1^{\nu+1}}\sum_{l=0}^\infty C_l(\nu)(\frac{u_2}{u_1})^l
\ee 
 the integral becomes
\ba\label{expansion}
&&I_{\nu_1,\nu_2,\nu_3} = - \oint \frac{du_1du_2du_3}{(2\pi i)^3}
\frac{(-1)^{\nu_1+\nu_2}[k_1k_2]^{\nu_1}[k_2k_3]^{\nu_2}[k_3k_1]^{\nu_3}}{u_1^{\nu_1+\nu_3+2}
u_2^{\nu_2+1}}
\nonumber\\
&& \times \sum_{l_i,s_j} C_{l_1}(\nu_1)C_{l_2}(\nu_2)C_{l_3}(\nu_3)(\frac{u_2}{u_1})^{l_1}
(\frac{u_3}{u_2})^{l_2}(\frac{u_3}{u_1})^{l_3}(i u_1)^{s_1}(i u_2)^{s_2}(i u_3)^{s_3}
\nonumber\\
&& J_{s_1}(2 t_1)J_{s_2}(2 t_2)J_{s_3}(2 t_3)
\ea
The u-integrals are now easy, and the coefficient of $(i t_1)^{k_1}(i t_2)^{k_2}(i t_3)^{2k_3}$,  denoted
 $I_{\nu_1,\nu_2,\nu_3}^{2k_1,2k_2,2k_3}$, is expressed as a sum.
This sum is given by the contour integral
\ba
&&I_{\nu_1,\nu_2,\nu_3}^{2k_1,2k_2,2k_3}= - (-1)^{\nu_1+\nu_2}[k_1k_2]^{\nu_1}[k_2k_3]^{\nu_2}[k_3k_1]^{\nu_3}
\nonumber\\
&&\frac{1}{\prod (2k_i)!}\oint \frac{dx dy dz}{(2\pi i)^3}
\sum (xy)^{\frac{l_1}{2}}(x z)^{\frac{l_3}{2}}(\frac{z}{y})^{\frac{l_2}{2}}C_{l_1}(\nu_1)C_{l_2}(\nu_2)
C_{l_3}(\nu_3)
\ea
We consider separately  (i) $l_1,l_2$  even, and $l_3$ odd, and (ii)$l_1,l_3$ even, and $l_2$  odd.
When we consider the universal scaling limit for large $k_i$, this difference can be neglected. 
We replace $l_i\to 2 l_i$ or $l_i\to 2l_i+1$. The sum
over $l_i$ becomes
\be
\sum_l C_{2l}(\nu)(x y)^{l} \sim \sum_l C_{2l+1}(\nu) (x y)^{l} \sim \frac{2^\nu}{(1- x y)^{\nu+1}}
\ee
For instance when $\nu=3$, we have
\ba
\sum_l C_{2l}(3) (x y)^l &=& \frac{1}{6}\sum (2l+3)(2l+2)(2l+1) (x y)^l\nonumber\\
&=& \frac{1 + 6 x y + (x y)^2}{(1 - x y)^4}
\ea
and, since the saddle point  is $x_c=y_c=1$, the numerator at $x=y=1$ is indeed equal to $2^3$ .

Then one has
\ba\label{3genus}
&&I_{\nu_1,\nu_2,\nu_3}^{2k_1,2k_2,2k_3}= - (-1)^{\nu_1+\nu_2}[k_1k_2]^{\nu_1}[k_2k_3]^{\nu_2}[k_3k_1]^{\nu_3}
\nonumber\\
&&\times \frac{1}{\prod (2k_i)!}
\oint \frac{dx dy dz}{(2 \pi i)^3}
\frac{(1+x)^{2k_1}(1+y)^{2k_2}(1 + z)^{2k_3} 2^{\nu_1+\nu_2+\nu_3}}{x^{k_1- \frac{\nu_1}{2}-\frac{\nu_3}{2}} 
y^{k_2+\frac{\nu_2}{2}+1} z^{k_3} (1 - x y)^{\nu_1+1}(1 - \frac{z}{y})^{\nu_2+1}(1 - x z)^{\nu_3+1}}
\nonumber\\
&&= \frac{1}{\prod (2k_i)!}\oint
\frac{dx dy dz}{(2 \pi i)^3}
\frac{(-1)^{\nu_3}(x+y)^{2k_1}(1+y)^{2k_2}(1 +  y z)^{2k_3} 2^{\nu_1+\nu_2+\nu_3}}{
x^{k_1} 
y^{k_1+k_2+k_3+1} z^{k_3} (x-1)^{\nu_1+1}(z -1)^{\nu_2+1}(x z-1)^{\nu_3+1}}\nonumber\\
&&\times [k_1k_2]^{\nu_1}[k_2k_3]^{\nu_2}[k_3k_1]^{\nu_3}
\ea
In the last line, we have changed  $x\to \frac{x}{y}$ and $z\to z y$. We have also dropped
the subleading powers of $x$ and $y$ for  large $k_i$.

We have from (\ref{Uvertices}),
\be\label{3tr}
\frac{1}{N^n} <\prod_{j=1}^3 {\tr} M^{2 k_j}>_c = \prod_{j=1}^3 (2 k_j)! (\sum_{\nu} I_{\nu_1,\nu_2,\nu_3}^{2k_1,2k_2,2k_3})
e^{\frac{1}{12N^2}\sum_{j=1}^3 k_j^3}
\ee

As discussed before, we need the term of order of $\frac{1}{N^4}$, and the
order $k^{3/2}$ in the large $k$ limit. For this reason, we take $\nu_1+ \nu_2+ \nu_3=1$ in (\ref{3genus}).

We have
\be
\frac{1}{(z-1)(x z-1)} = \frac{1}{z(x-1)}(\frac{1}{z-1} - \frac{1}{x z - 1})
\ee
Using this identity, we have for $\nu_1+\nu_2+\nu_3=1$,
\ba
&&\frac{J}{N}= \frac{[k_1k_2]}{(x-1)^2 (z-1)(x z-1)} 
+ \frac{[k_2 k_3]}{(x-1)(z-1)^2(x z-1)}\nonumber\\
&&+ \frac{[k_3 k_1]}{(x-1)(z-1)(x z-1)^2}\nonumber\\
&&=- \frac{k_3}{N  s_1^3  s_3} + \frac{k_2+k_3}{2N s_1^2  s_3^2}
+ \frac{k_3}{N s_1^3 ( s_1 +  s_3)}+ \frac{k_3+k_1}{2 N  s_1^2 ( s_1+  s_3)^2}
\ea
where we have  expanded $x,y$ and $z$ near the saddle points as $x= 1 + i s_1,y=1+ i s_2, z=1+ i s_3$.
From  the saddle point analysis, we obtain
\ba
&&\frac{1}{N} \int \prod_{j=1}^3 \frac{d s_j}{(2\pi i)} e^{-\frac{k_1}{4} s_1^2 - \frac{k_1+k_2+k_3}{4}
s_2^2 - \frac{k_3}{4} s_3^2 + \frac{k_1}{2} s_1 s_2 + \frac{k_3}{2} s_2 s_3}
J\nonumber\\
&&= I_1 + I_2
\ea
where $I_1$ corresponds to the first two terms of $J$, and $I_2$ corresponds to 
the third and fourth.
For the  $s$-integral, we integrate by parts, 
which reduces the integrand
of $J$ to the sum of  constant terms and single pole terms in  $s_1$ and $s_3$.
We take only the constant terms after diagonalization of the quadratic form in the  exponent.
We obtain
\ba
&&I_1 = \frac{\sqrt{k_1 k_2 k_3}(k_2+k_3)}{8 N \pi^{3/2}(k_1+k_2+k_3)}\nonumber\\
&&I_2=  \frac{\sqrt{k_1 k_2 k_3} (k_1+k_3)}{8 N \pi^{3/2}(k_1+k_2+k_3)}
\ea
Returning to the expansion of (\ref{expansion}), 
we add the  permutations over the $k_i$,  and obtain
\be
\frac{1}{N^3}<{\tr} M^{2k_1}{\tr}M^{2 k_2} {\tr}M^{2 k_3}>_c
= \frac{4^{k_1+k_2+k_3}}{N^4} \frac{\sqrt{k_1 k_2 k_3}}{\pi^{3/2}}
\ee
Dividing by  by $\frac{4^{k_1+k_2+k_3}\sqrt{k_1 k_2 k_3}}{\pi^{3/2}}$,  we obtain the intersection number 
\be
< \tau_0^3 >_{g=0 }= 1
\ee
The intersection numbers, for the three point $n=3$, for  higher genuses
 are evaluated by considering higher values 
 for $\nu_i$ in (\ref{3tr}).
After factoring out $\frac{4^{k_1+k_2+k_3}\sqrt{k_1 k_2 k_3}}{\pi^{3/2}}$, 
the vertex correlations $\frac{1}{N^3}<{\tr} M^{2k_1}{\tr}M^{2 k_2} {\tr}M^{2 k_3}>_c$
become polynomials in $k_i$, and scale as $\frac{k^3}{N^2}$. The intersection numbers
are obtained from the coefficients of these polynomials.

\vskip 2mm
\section{Characteristic polynomials for  Airy matrix functions and the replica method}
\vskip 2mm

We consider in this section  the
correlation functions for the characteristic polynomials of  random matrices. The Airy
matrix model of Kontsevich type,  will then be derived from these correlation functions at
the edge of the spectrum.

In a recent work we have studied  the average of  products of  characteristic
polynomials 
 \cite{BHe},  defined as 
\ba
F_k(\lambda_1,...,\lambda_k) &=& < \prod_{\alpha=1}^k {\rm det}(\lambda_\alpha - M) >_{A,M}\nonumber\\
&=& \int dM \prod_{i=1}^k {\rm det}(\lambda_i\cdot {\rm I} - {\rm M}) e^{-\frac{N}{2}
\tr M^2 + N  \tr M A} 
\ea
where $M$ is an $N\times N$ Hermitian random matrix.
It was shown that this correlation function has also a dual expression.  This duality interchanges $N$, the
size of the random matrix,  with
$k$, the number of points in $F_k$,   as well as the matrix source $A$ with the diagonal
matrix  $\Lambda = {\rm diag}(\lambda_1,...,\lambda_k)$. Indeed we had derived that this same correlation function is given by 
\cite{BHe}
\be\label{dualB}
F_k(\lambda_1,...,\lambda_k)= \int dB \prod_{j=1}^N [{\rm det}(a_j - i B)] e^{-\frac{N}{2}\tr B^2 +
i N \tr B \Lambda}
\ee
where $\Lambda = {\rm diag}(\lambda_1,...,\lambda_k)$ and $B$ is a $k \times k$ 
Hermitian matrix.

If we specialize this formula to a source $A$ equal to the unit matrix,  providing thus a
trivial  constant shift for
$M$, the formula (\ref{dualB}) involves 
\ba\label{expandB}
{\rm det}(1 - i B)^N &=& {\rm exp}[ N {\tr} {\rm ln}(1 - i B)] 
\nonumber\\
&=&{\rm exp}[ - i N {\tr} B + \frac{N}{2} {\tr} B^2 + i \frac{N}{3}{\tr} B^3 + \cdots]
\ea

The linear term in $B$ in (\ref{expandB}), combined with the linear term of the exponent
of (\ref{dualB}), shifts $\Lambda$ by one.
The $B^2$ terms in (\ref{dualB}) cancel. In a scale in which the initials $\lambda_k$
are close to one, or more precisely $N^{2/3}(\lambda_k-1)$ is finite, the large $N$
asymptotics of (\ref{dualB}) is given by matrices $B$  of order
$N^{-1/3}$. Then the higher terms in (\ref{expandB}) are negligible and we are left with
terms linear and cubic in the exponent, namely
\be\label{dualB2}
F_k(\lambda_1,...,\lambda_k)= \int dB \prod_{j=1}^N  e^{i\frac{N}{3}{\tr} B^3 +
i N {\tr} B (\Lambda-1)}.
\ee  This is nearly identical to 
the matrix Airy integral, namely Kontsevich's model \cite{Kontsevich}, which gives  the
intersection numbers of moduli of curves.

The original Kontsevich partition function was defined as 
\be
Z = \frac{1}{Z^\prime}\int dM e^{-\frac{1}{2}\tr \Lambda M^2 + \frac{i}{6}\tr M^3}
\ee
where $Z^\prime = \int dM e^{-\frac{1}{2}\tr M^2}$. The shift $M \to M - i \Lambda$,
eliminates the ${M^2}$ term and one recovers (\ref{dualB2}).

 Let us examine  the simple case of a  $2\times 2$ matrix $M$, with two points
$\lambda_1$ and $\lambda_2$. (Indeed this simple $N=2$ case is useful as a check for the
intersection numbers. For higher N
one could perform a similar analysis.) Then we have
\be
Z = \sqrt{\lambda_1 \lambda_2}(\lambda_1 + \lambda_2)e^{\frac{1}{2}(\lambda_1^3 + \lambda_2^3)}
Y
\ee
where
\ba
Y &=&\int dM e^{\frac{i}{2}\tr \Lambda^2 M + \frac{i}{6}\tr M^3}\nonumber\\
&=&\int dx_1 dx_2 (\frac{x_1 - x_2}{\lambda_1^2 - \lambda_2^2})
e^{\frac{i}{2}(\lambda_1^2 x_1 + \lambda_2^2 x_2)
+ \frac{i}{6}(x_1^3 + x_2^3)}\nonumber\\
&=&\frac{2}{\lambda_1^2 -\lambda_2^2}(\frac{\partial}{\partial \lambda_1^2} - 
\frac{\partial}{\partial \lambda_2^2})[\frac{1}{\sqrt{\lambda_1 \lambda_2}}e^{-\frac{1}{3}
(\lambda_1^3 + \lambda_2^3)}z(\lambda_1)z(\lambda_2)]
\ea
where 
\be
z(\lambda) =\frac
{\int dx e^{-\frac{x^2}{2}\lambda + \frac{i}{6}x^3}}{\int e^{-\frac{x^2}{2}
\lambda} dx}
\ee
Then we get
\be
Z=1 + \frac{1}{6} \tilde t_0^3 + \frac{1}{24}\tilde t_1
 + O(\frac{1}{\lambda^5}),
\ee
\ba\label{logZ}
&&\log Z = \frac{1}{6}\tilde t_0^3 + \frac{1}{24}\tilde t_1 + \frac{1}{6}\tilde t_0^3 \tilde t_1
+\frac{1}{48}\tilde t_1^2 + \frac{1}{24}\tilde t_0 \tilde t_2 \nonumber\\
&&+ \frac{1}{6}\tilde t_1^2 \tilde t_0^3 + \frac{1}{72}\tilde t_1^3 + \frac{1}{48}\tilde t_3 t_0^2
+ \frac{1}{12}\tilde t_0 \tilde t_1 \tilde t_2 \nonumber\\
&& + \frac{1}{1152}\tilde t_4 + O(\frac{1}{\lambda^{12}})
\ea
where we have defined the moduli parameter $\tilde t_i$ as
\be
\tilde t_0=\frac{1}{\lambda_1} + \frac{1}{\lambda_2}, \hskip 3mm
\tilde t_1 = \frac{1}{\lambda_1^3} + \frac{1}{\lambda_2^3}, \hskip 3mm
\tilde t_n =  \frac{1}{\lambda_1^{2n+1}}+ \frac{1}{\lambda_2^{2n+1}}.
\ee
(We need the notation $\tilde t_n$, instead of the usual notation $t_n$, 
  for distinguishing those parameters from the Fourier transform 
parameters $t_i$).

From these coefficients, we recover the known results 
\ba
&&<\tau_0^3>_{g=0}= 1, \hskip 4mm <\tau_1>_{g=1}=\frac{1}{24}, \hskip 4mm
<\tau_0^3 \tau_1>_{g=0} = 1, \nonumber\\
&&< \tau_1^2>_{g=1} = \frac{1}{24},\hskip 4mm <\tau_0 \tau_2>_{g=1} =\frac{1}{24}\nonumber\\
&& <\tau_0^3 \tau_1^2>_{g=0}= 2, \hskip 4mm <\tau_1^3>_{g=1} = \frac{1}{12},
\hskip 4mm <\tau_0^2 \tau_3>_{g=1} = \frac{1}{24}\nonumber\\
&&<\tau_0 \tau_1 \tau_2>{g=1}= \frac{1}{12},\hskip 4mm <\tau_4>_{g=2} = \frac{1}{1152}
\ea

The generating function for the intersection numbers $<\prod \tau_i^{d_i}>_g$ is
\be\label{generating2}
\log Z = \sum_{m_l} <\tau_0^{m_1}\tau_1^{m_2}\cdots> \prod_{l=0}^\infty
\frac{\tilde t_l^{m_l}}{
m_l!}
\ee
where $\tilde t_l$ is related to the eigenvalues of the matrix $\Lambda$ as
\be
\tilde t_l = \sum_{j=1}^N \frac{(2l-1)!!}{\lambda_j^{2l+1}}
\ee

Note that the generating function of the intersection numbers  (\ref{generating2})
is quite different from the generating function  (\ref{generating1}), which was obtained 
from
the correlation functions of the vertices. The relation between $\tilde t_l$ and $x_j$ are
\be
\tilde t_l \sim x_j^l
\ee
where $x_j = \frac{k_j}{2^{1/3}}$, in which $k_j$ is the power in  $M^{2k_j}$.
However the degenerate case needs some precaution, when we have $\tilde t_l^2 \sim x_j^l
x_m^l$ for $j\ne m$.


The replica method for the correlation functions has been used earlier in  random
matrix theory for the GUE
\cite{BHj,BHg}. Following this replica analysis  we
study now two types of correlation functions.  The first one is the  correlation function
for the eigenvalues of the matrix $M$ :
\be\label{correlation1}
\rho(\lambda_1,...,\lambda_k) = <\prod_{\alpha=1}^k \frac{1}{N}{\tr} \delta(\lambda_\alpha - M)>_{A,M}
\ee
 The second type of  correlation functions is the average of 
products of  characteristic polynomials \cite{BHe},
\be\label{correlation2}
 F_k(\lambda_1,...,\lambda_k)=<\prod_{\alpha=1}^k {\rm det}(\lambda_\alpha - M)>_{A,M}. 
\ee
where the average $<...>_{A,M}$ is with respect to  the probability distribution $P_A(M)$,
\be
P_A(M) = \frac{1}{Z}e^{-\frac{N}{2}{\tr} M^2 + N {\tr} M A}
\ee
The random matrix $M$ is a complex Hermitian N by N matrix and $A$ is an external source,
which we can take as a diagonal matrix, $A= {\rm diag} (a_1,...,a_N)$ since the integration measure is unitary invariant. If $A$ is zero, it reduces
to  the Gaussian unitary ensemble (GUE), but  it is convenient to use the
probability distribution
$P_A(M)$ for setting up the replica method, even if we let $A=0$ at the end. The correlation
functions with the distribution 
$P_A(M)$ require the HarishChandra-Itzykson-Zuber formula
\cite{Harish-Chandra,Itzykson-Zuber} for the unitary group integral.

We have evaluated the F.T. of the correlation functions near the edge
in the previous sections.
To prove that F.T. of the correlation function is the generating function
of the intersection numbers,
we express these functions as the zero replica limit of characteristic polynomials
 average.
Let us begin with the one point function, namely the density of states ; we first use the
identity
\be\label{r1}
< {\tr} \delta (\lambda - M) >_{A,M} =\frac{1}{\pi}{\Im}{\rm m}\hskip 1mm
\lim_{ n\to 0} \frac{1}{n} \frac{\partial}{\partial \lambda} 
< [{\rm det}(\lambda -i\epsilon- M)]^n >_{A,M}\ee
and use the duality derived in \cite{BHe} to write
\be\label{r2} < [{\rm det}(\lambda - M)]^n >_{A,M} =< \prod_{\gamma=1}^N  [{\rm
det}(a_\gamma - i B)] >_{\Lambda,B}
\ee
where $B$ is an $n\times n$ random Hermitian matrix, and $\Lambda $, in this case,   is a
multiple of the
$n\times n$ identity  matrix :
$ \Lambda = {\rm diag}(\lambda,...,\lambda)$. Note that we have traded an  $n$-point
function of 
$N\times N$  matrices $M$ for an $N$ point function  of $n\times n$  matrices $B$.

If we explore the edge of the distribution by taking again $a_\gamma=1$, and expand
in powers of
$B$, we find that the
$B^2$ term cancels, and obtain
\be
U(t) =  <\frac{1}{N} {\tr} \hskip 1mm e^{i t M} > =\Im {\rm m} \hskip 1mm \lim_{n\to 0}
\int d\lambda e^{i t \lambda}  
 \frac{1}{n} \frac{\partial}{\partial \lambda}
\int dB e^{ {\tr} [i \frac{N}{2}B^3 + i B (\lambda-1)]}
\ee
After integration by parts over $\lambda$ we obtain 
\be\label{reU}
U(t) = t \hskip 1mm \rm \lim_{n\to 0} \int d\lambda e^{i t \lambda}  \frac{1}{n} 
\int dB e^{ i {\tr}[\frac{N}{2}B^3 + i B (\lambda-1)]}
\ee
Since the replica parameter $n$ means the repetition $n$ times of the same $\lambda$ ,
we simply replace the $\tilde t_l$ of the previous un-replicated case by

\be
\tilde t_l=\sum_{j=1}^k \frac{1}{\lambda_j^l} \to \sum_{j=1}^k \frac{n}{\lambda_j^l}
\ee
For the one point function k=1,
the expansion of the Airy matrix model in terms of $\tilde t_l$ is thus a power series in
$n$. In the zero-replica limit one can neglect all  terms beyond the linear one in $n$.
Thus, $U(t)$ in (\ref{reU})
is expressed as  linear combinations of the $\tilde t_l$. The F.T. of $\tilde t_l =
\frac{1}{\lambda^l}$ yields dimensionally  a factor
$t^l$. Therefore,
$U(t)$ appears as a power series in $t$, whose coefficients are the intersection
numbers 
$<\tau_l>$, the same coefficients  obtained in  the Kontsevich model.

This argument holds also for the two point correlation function, since
\be\label{r2}
\lim_{n\to 0}\frac{\partial^2}{\partial \lambda_1 \partial \lambda_2}\frac{1}{n^2}
<[{\rm det}(\lambda_1 - M){\rm det}(\lambda_2 - M)]^n>
= \tr \frac{1}{\lambda_1 - M} \tr \frac{1}{\lambda_2 - M}
\ee

In this two-point case (k=2), one deals with $\lambda_1$ and $\lambda_2$ and $\tilde t_l$
becomes
\be
\tilde t_l = \frac{n}{\lambda_1^l} + \frac{n}{\lambda_2^l}
\ee
The replica limit ($n \to 0$) requires to retain the terms of order $n^2$ ,
i.e. products of the form $\tilde t_l \tilde t_m$.
Then  $U(t_1,t_2)$ is a power series in $t_1,t_2$, whose coefficients are
 the intersection numbers
$<\tau_l \tau_m>$.

 The F.T. of the k-point correlation function $U(t_1,...,t_k)$ produces, by the same
replica method, the intersection numbers $<\tau_{m_1}\tau_{m_2}...>$ as coefficients
of
$t_1^{m_1}t_2^{m_2}\cdots t_k^{m_k}$.

Thus we have shown that the F.T. of the correlation functions in the edge region, is the
generating function of the intersection numbers , our main conclusion.


\section{Summary}

We have derived the expressions for the correlation functions of the vertices
$<\frac{1}{N^n}\prod_{i=1}^n{\tr} M^{2k_i}>$, when
$N$ and $k_i$ are simultaneously large, in a scaling region in which $k^3$ scales like $N^2$. The expressions are obtained from an exact  contour integral representation valid for finite N and for an arbitrary external matrix source.
The coefficients of the expansion of the correlation functions   provide
the intersection numbers of the moduli space of curves.

The correlation functions of the eigenvalues may also be obtained from the average of characteristic polynomials in a zero-replica limit. Using a previously derived duality in which the size N of the matrix is interchanged with the number of points of the correlation functions, we have  recovered the Airy matrix model of Kontsevich and re-derived the intersection numbers by a simple saddle-point analysis.


\end{document}